\documentclass[journal=apchd5,manuscript=article,layout=traditional]{achemso}

\usepackage{graphicx}
\usepackage{amsmath}
\usepackage{amssymb}
\usepackage{booktabs}
\usepackage[version=4]{mhchem}
\usepackage{pdfpages}

\graphicspath{{figures/}}

\SectionNumbersOff

\newcommand{\dE}{\ensuremath{\Delta E_{00}}}

\author{Zhiyi Li}
\affiliation[xmu]{School of Electronic Science and Engineering (National Model Microelectronics College), Xiamen University, Xiamen, Fujian 361005, China}
\alsoaffiliation[fjkl]{Fujian Key Laboratory of Ultrafast Laser Technology and Applications, Xiamen University, Xiamen 361005, China}
\author{Yuheng Jin}
\affiliation[xmu]{School of Electronic Science and Engineering (National Model Microelectronics College), Xiamen University, Xiamen, Fujian 361005, China}
\alsoaffiliation[fjkl]{Fujian Key Laboratory of Ultrafast Laser Technology and Applications, Xiamen University, Xiamen 361005, China}
\author{Yidan Huang}
\affiliation[xmu]{School of Electronic Science and Engineering (National Model Microelectronics College), Xiamen University, Xiamen, Fujian 361005, China}
\alsoaffiliation[fjkl]{Fujian Key Laboratory of Ultrafast Laser Technology and Applications, Xiamen University, Xiamen 361005, China}
\author{Nan Chen}
\affiliation[xmu]{School of Electronic Science and Engineering (National Model Microelectronics College), Xiamen University, Xiamen, Fujian 361005, China}
\alsoaffiliation[fjkl]{Fujian Key Laboratory of Ultrafast Laser Technology and Applications, Xiamen University, Xiamen 361005, China}
\author{Hongyan Fu}
\email{fuhongyan@xmu.edu.cn}
\affiliation[xmu]{School of Electronic Science and Engineering (National Model Microelectronics College), Xiamen University, Xiamen, Fujian 361005, China}
\alsoaffiliation[fjkl]{Fujian Key Laboratory of Ultrafast Laser Technology and Applications, Xiamen University, Xiamen 361005, China}
\author{Yikun Bu}
\email{buyikun0522@xmu.edu.cn}
\affiliation[xmu]{School of Electronic Science and Engineering (National Model Microelectronics College), Xiamen University, Xiamen, Fujian 361005, China}
\alsoaffiliation[fjkl]{Fujian Key Laboratory of Ultrafast Laser Technology and Applications, Xiamen University, Xiamen 361005, China}

\title{Flexible Inverse Design of Optical Multilayer Coatings Using
Discrete--Continuous Flow Matching}

\keywords{optical thin films, multilayer coatings, inverse design, generative
models, flow matching, structural color}

\begin{document}

\begin{abstract}
Optical multilayer coatings are used in imaging optics, lasers,
telecommunications, solar cells, and radiative cooling, where the spectral
response required by each application is achieved by a well-designed layer
structure. However, existing design methods have limitations. Local methods refine a starting design efficiently but
depend on its quality, and global searches require
up to millions of transfer-matrix evaluations per target. Existing
deep-learning inverse-design models are fast, but they are trained for a fixed
set of materials and a fixed wavelength band, and they either optimize the
thicknesses for a fixed material sequence or choose them from a discrete grid.
In this paper, we present IrisFlow, a generative model that addresses these
limitations by taking each candidate material as its optical constants and the target spectrum together with its
wavelengths as input. It generates the material and the continuous thickness of
every layer jointly, by coupled discrete and continuous flow matching in a
single network. We demonstrate its performance through design with materials
outside the training library and on bands beyond the training range, and
through comparisons with classical optimizers and an existing deep-learning
model. We also demonstrate its usability in a
practical design workflow by designing four colored solar-reflective coatings with
the calibrated optical constants of an existing deposition system. Combined with a local refinement of its ten best
designs, it reduces the number of transfer-matrix evaluations about 500-fold
relative to needle optimization at comparable accuracy. The code, trained model, and training and test data are publicly available.
\end{abstract}

\section{Introduction}

Optical thin films find broad application in imaging lenses,\cite{dobrowolski2002}
consumer electronics, solar cells,\cite{bouhafs1998} lasers and
spectrometers, telecommunications,\cite{macleod2010} and
aerospace. Beyond these established uses, they have more recently been employed
for passive radiative cooling\cite{raman2014,zhai2017} and for
selective shading in greenhouse agriculture.\cite{shi2019,li2024}
Designing a coating means choosing a discrete material and a continuous
thickness for every layer. The optical constants of each material depend on the deposition process
and conditions, so each
coating facility maintains its own calibrated dispersion data and updates them
as processes change. Among conventional optimization methods, local refinement and synthesis methods such as needle
optimization, gradual evolution, and gradient-based optimization refine a
reasonable starting design to high
precision,\cite{tikhonravov1996,sullivan1996,tikhonravov2007} while global
searches such as genetic algorithms, particle swarm optimization, and simulated annealing reduce the dependence on the initial
guess.\cite{martin1995,rabady2014,chang1990} Both families, however, start a new search for every target. Local methods
are fast but easily become
trapped in local optima, and global methods are more robust but require
thousands to millions of forward evaluations with the transfer-matrix method (TMM).

\begin{figure}[htbp]
  \centering
  \includegraphics[width=\linewidth]{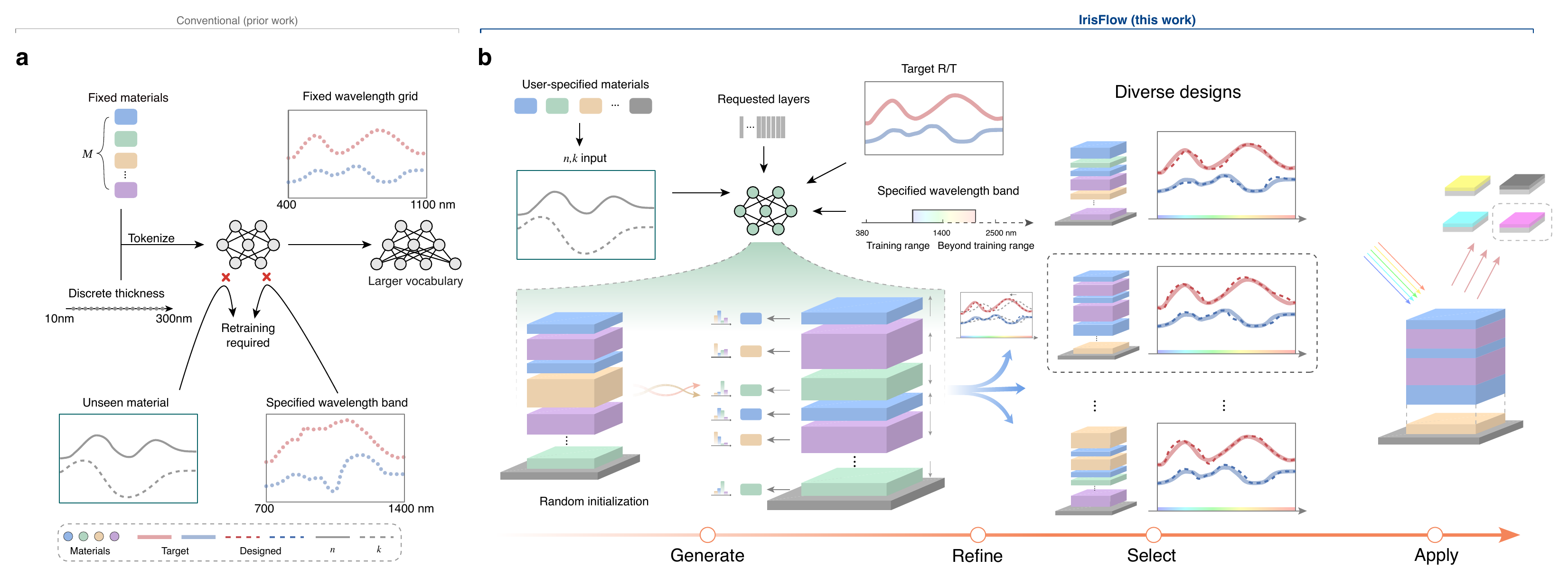}
  \caption{IrisFlow: a generative framework for inverse design of multilayer
  thin films with user-specified materials, layer counts, and wavelength
  bands. (a) Existing machine-learning models: materials are encoded as discrete
  tokens from a fixed library, thicknesses take values on a fixed grid, and target spectra are defined on the band used in training; a new material or a
  new band requires retraining. (b) IrisFlow: materials enter through their optical constants $n(\lambda)$ and $k(\lambda)$, the layer count is specified by
  the user, and the target spectrum is supplied together with its wavelength
  band, which may extend beyond the training range; the materials and the
  continuous thicknesses of all layers are generated jointly by coupled
  discrete and continuous flows.}
  \label{fgr:overview}
\end{figure}

Machine learning can reduce the number of TMM evaluations, avoid local optima,
and find better designs by learning the inverse mapping from spectrum to
stack.\cite{molesky2018,ma2021} For multilayer coatings,\cite{ma2025} a trained
network proposes designs for a new target without running a new search. An early demonstration used a fully connected network as a surrogate for
the optical response of multilayer core--shell nanoparticles and inverted it by
gradient descent on its inputs.\cite{peurifoy2018} Tandem networks train an
inverse model through a fixed forward predictor to address the non-uniqueness
of the inverse problem,\cite{liu2018} and mixture-density networks represent
multiple admissible thickness solutions as conditional probability
distributions.\cite{unni2020,unni2021} Once trained, these models perform
inverse design from a target spectrum, but only as thickness optimization for a
given material sequence. Reinforcement learning instead treats the stack as a
sequential decision process and can choose materials as well as thicknesses,
but it must run a new search for each target.\cite{jiang2020,wankerl2021} OptoGPT removes the fixed-structure restriction by serializing a
structure as a sequence of discrete material and thickness tokens and
generating both with a decoder-only transformer;\cite{ma2024optogpt} a masked
diffusion decoder has since been applied to the same
formulation.\cite{schaible2026} This formulation, however, has three
limitations. First, the material vocabulary is tied to the optical
constants used to generate the training data, so at inference the model can
only choose from that vocabulary; whenever the optical constants are updated or
a new material is introduced, the training data must be regenerated and the model retrained. Second, thickness discretization imposes a
trade-off: coarse steps sacrifice the precision on which thin-film performance
depends and can miss the optimum, whereas fine steps require a larger
vocabulary, more training data, and correspondingly longer training. Third, the target spectrum must be supplied on the wavelength band used in
training, so the model cannot be applied directly when
the operating band changes, as illustrated in Figure~\ref{fgr:overview}a. A practical design workflow
needs a model that adapts to changing optical constants and wavelength bands
without retraining and keeps the thicknesses continuous.

In this work, we introduce IrisFlow, a generative model for multilayer coating
design that addresses these three limitations (Figure~\ref{fgr:overview}).
Instead of drawing materials from a fixed vocabulary, IrisFlow receives each
candidate material as its optical constants $n(\lambda)$ and $k(\lambda)$, so
that updated optical constants or new materials can be supplied at inference
without retraining. Furthermore, the model couples discrete material
generation and continuous thickness generation by joint discrete and
continuous flow matching\cite{lipman2023,campbell2024,gat2024} in a single
network, which avoids both the restriction to optimizing thicknesses for a
fixed material sequence and the trade-off imposed by discretized thicknesses.
Finally, the target spectrum is provided together with its wavelengths, so
the model is not tied to the training band and can design for bands beyond the
wavelength range seen in training. The designs sampled for a target are structurally diverse, so the designer can
select among them according to further requirements, such as ease of
deposition. Moreover, the generated designs can serve as starting points for
conventional methods such as local refinement.

We demonstrate the performance of IrisFlow through several tests, including
spectral reconstruction within the training distribution, transfer to
materials outside the training library, design for wavelength bands beyond
the training range, and design for idealized target spectra outside the
training distribution. We
further compare IrisFlow with classical numerical optimizers, and with OptoGPT
using the optical constants reported in the OptoGPT study. We then show design under constraints set by the designer and at oblique
incidence.
Finally, we use IrisFlow in a practical design workflow to design four colored
solar-reflective coatings with the calibrated optical constants of an existing
deposition system, from sampling through refinement toward the target color and
selection to the tolerance of the designs to thickness errors.

\section{Methods}

\subsection{Physical input representation}

\begin{figure}[htbp]
  \centering
  \includegraphics[width=\linewidth]{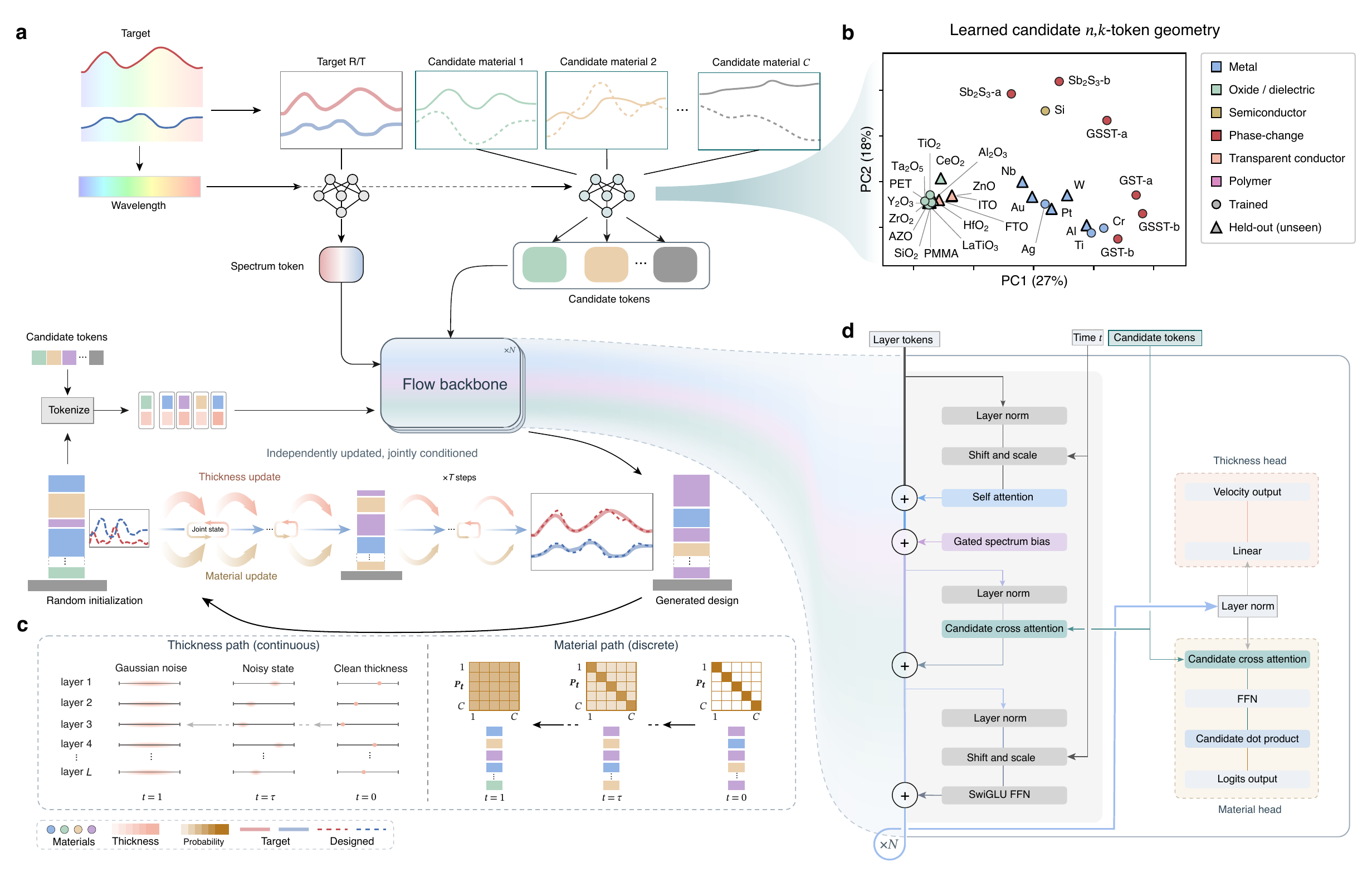}
  \caption{Architecture and generative process of IrisFlow. (a) Input
  representation and sampling loop. The target $R$ and $T$ spectrum and the optical constants of the $C$ candidate materials, each supplied together with its
  wavelength grid, are encoded by two multilayer perceptrons of the same architecture into one
  spectrum token and $C$ candidate tokens; layer tokens are formed by tokenizing the current material and
  thickness of every layer. At each step, the thickness and material updates
  of every layer are computed from the same layer tokens, which carry the
  current thicknesses and materials of the whole stack.
  (b) First two principal components of the learned candidate-material tokens
  (27\% and 18\% of the variance), for the 15 training materials (circles) and
  the 15 held-out materials of Figure S5 (triangles), colored by material class. (c) The two noising paths. Thickness path (continuous): Gaussian
  noise at $t=1$, the noisy state at an intermediate $t=\tau$, and the clean
  thickness at $t=0$ for each of the $L$ layers. Material path (discrete): the
  transition matrix $\mathbf{P}_t$ over the $C$ candidates, uniform at $t=1$
  and the identity at $t=0$, with the corresponding material state below.
  (d) Structure of one network block.}
  \label{fgr:model}
\end{figure}

To address the fixed material vocabulary and the fixed wavelength band of existing machine-learning models, we represent each candidate material by its optical constants
$n(\lambda)$ and $k(\lambda)$ and the target by its spectrum $R(\lambda)$ and
$T(\lambda)$, each supplied together with the wavelengths at which it is
tabulated. Wavelength is encoded as a feature linear in $1/\lambda$ (Note S3).
The optical constants and the spectrum are two-channel functions of wavelength
and are encoded by two
multilayer perceptrons (MLPs) of the same architecture: one produces a spectrum token, the other one candidate token
per material (Figure~\ref{fgr:model}a). A material therefore enters
the model only through its optical constants, so updated optical constants, or
materials outside the training library, require no change to the model.

After training, we applied principal component analysis (PCA) to the candidate tokens of the 15 training materials and projected the 15
held-out materials (Figure S5) into the same space. As shown in Figure~\ref{fgr:model}b, the learned tokens are arranged in
order of material class, and the held-out metals and oxides are placed among
their training counterparts, although the model never received a material
label. Figure S7 shows the analysis for the raw optical constants.

\subsection{Discrete--continuous flow matching}

Inverse design of a multilayer coating is a one-to-many problem: many
different stacks give nearly the same spectrum, so the designs that meet a
target form a multimodal distribution. We therefore build IrisFlow on flow
matching,\cite{lipman2023,campbell2024,gat2024} a class of generative models
that learn to transform noise into samples of a data distribution along a
prescribed path and can represent such multimodal distributions. IrisFlow
learns how to transform a random stack into a design, conditioned on the target spectrum and
the candidate materials. The thicknesses and the materials are generated
jointly by one network (Figure~\ref{fgr:model}d); only their noising paths
differ. As shown
in Figure~\ref{fgr:model}c, the thicknesses follow the
linear flow-matching path from their clean values at $t=0$ to standard Gaussian
noise at $t=1$; the materials follow a continuous-time Markov chain (CTMC) with
a uniform kernel over the candidate set supplied for the task, which takes the
clean sequence at $t=0$ to a uniform distribution over the candidates at $t=1$
(Note S1).

Each layer is represented by a layer token that combines the token of its
current material with an encoding of its current thickness. The layer tokens
attend to one another and to the candidate tokens and are conditioned on the
spectrum token (Figure~\ref{fgr:model}d; Note S3). From these tokens, a
continuous head predicts the thickness velocity, and a discrete head scores
each layer token against the candidate tokens to give, for each layer, a
probability distribution over the supplied candidates. Because both heads read the same tokens, each prediction is
conditioned on the current material and thickness of every layer, which
couples the discrete and the continuous paths. The number of layers is specified by the user, and the positions beyond it,
up to the maximum of 100, are masked.

\subsection{Dataset}

Training spectra are computed with the TMM\cite{macleod2010} at normal
incidence over 380--1400~nm, with air as the incident medium and a 1~mm
fused-silica substrate whose back surface is included as an incoherent
reflection. The target spectrum of each sample covers one band or two separated bands
within this range, and this spectrum and the optical constants of the
candidates share one 128-point grid, uniform in $1/\lambda$. The training
library contains 15 materials (Table S1). Layer thicknesses lie within
5--300~nm, in windows that depend on the material. Each candidate set holds
the materials of the stack plus unused materials from the library, up to 15 in
total, in random order.

Most stacks of 2--20 layers are generated at random under rules that exclude
stacks that cannot be deposited (Table S2); the rest are multilayer dielectric
stacks (Note S2). Random stacks of more than 20 layers are
mostly opaque, so stacks of 21--100 layers are instead obtained by refining
random stacks toward target spectra. The training set for
2--20 layers contains 110 million samples; those for 2--60 and 2--100 layers
contain 1.6 and 2.5 million, of which 63\% and 76\% are refined long stacks
(Note S2).

\subsection{Training}

During training, the model predicts the thickness velocity and the material
probabilities from the noisy stack at time $t$, the candidate set, and the target spectrum. The thickness loss is the mean squared error between the predicted and the
target velocity, and the material loss is a cross-entropy on the clean
material, both over the layers of the stack. The training
loss is their weighted sum. Details are given in Note S1.

Training follows a curriculum. The model is first pretrained on short stacks of
2--20 layers and then extended to 100 layers in two further stages, 2--60 and
2--100 layers, each continuing from the previous stage and each retaining short stacks in its training set. The network is the same in all three stages; only the maximum number of
layers and the training set change. The detailed
training settings for each stage are given in Note S4, and the training progress
in Figure S2.

\subsection{Sampling}

After training, IrisFlow generates a design by integrating the joint
discrete--continuous flow in reverse, from noise at $t=1$ to a design at
$t=0$, updating the thicknesses and materials of all layers together at every
step (Figure~\ref{fgr:model}a; Note S1). The time steps are set
adaptively from the entropy of the material probabilities, and all reported results use 15 steps (Notes S1 and S5).

Because each run starts from different noise, repeated sampling gives
structurally different designs for the same target. Each target is sampled $N$
times, every design is re-simulated, and the one with the lowest error is
reported. We set $N=100$ for the in-distribution and held-out-material test sets. For the idealized targets, $N=500$ at each layer count of the sequence 1, 2, 4, 8, \ldots{} below the largest layer count specified for the target and at that count. Sampling for the colored coatings and for the comparison with classical optimizers is described in Notes S12 and S9.

Local constraints are imposed by resetting the specified materials and
thicknesses to their given values at every step, so that the final design
satisfies them exactly.

\subsection{Evaluation metric}

The generated stack is re-simulated with the TMM under the conditions of the
training data, on the wavelength grid of the target, and the spectral error is reported as the root-mean-square error (RMSE) over reflectance $R$ and transmittance $T$,
\begin{equation}
  \mathrm{RMSE}=\sqrt{\frac{1}{2S}\sum_{s=1}^{S}
  \Big[(\hat R_s-R^{*}_s)^2+(\hat T_s-T^{*}_s)^2\Big]},
  \label{eq:rmse}
\end{equation}
where $S$ is the number of wavelength points of the target, $\hat R,\hat T$ are
the re-simulated spectra and $R^{*},T^{*}$ the target.

For the colored coatings, color is compared on the CIE 1931 chromaticity
diagram, with the color difference given by CIEDE2000 (\dE{})\cite{sharma2005} under
D65 illumination with the CIE 1931 2$^\circ$ standard observer.

\subsection{Conversion for oblique incidence}

Because the model takes optical constants as input, an oblique-incidence design can be posed as a normal-incidence one by
supplying optical constants transformed to reproduce the tilted optical
admittance of each layer at the given angle and polarization\cite{macleod2010}
(Note S11).

\section{Results and Discussion}

\subsection{Inverse-design performance and generalization}

\subsubsection{Inside the training distribution}

We first tested the model on target spectra computed from stacks generated in
the same way as the training data. The test covered seven
layer-count ranges, from 2--5 to 81--100 layers, and six wavelength ranges,
five for single-band and one for dual-band targets, with 100 independent
targets for each combination; the band of each target was drawn at random
within its wavelength range, and the candidate set consisted of the 15
training materials.

As shown in Figure~\ref{fgr:general}a, the median RMSE changes little with the
target band: over all layer counts, it lies between 0.023 (1000--1400~nm) and
0.044 (380--550~nm) for the six wavelength ranges, and for every layer-count
range it varies by less than a factor of 2.5 across them. The median RMSE rises
with the number of layers, from 0.0077 at 2--5 layers to 0.068 at 81--100
layers (black line, Figure~\ref{fgr:general}d), likely because the number of possible
stacks grows exponentially with the layer count and stacks of more than 20
layers make up a much smaller share of the training data than short ones
(Note S2).
Figure~\ref{fgr:general}b,c shows two representative designs with RMSE close
to the median.

\subsubsection{Transfer to unseen materials}

We next tested transfer to materials that never appeared in training. We
replaced the candidate set with 15 held-out materials (Figure S5) and repeated
the same test. As shown in Figure~\ref{fgr:general}d, the error with the
held-out materials is of the same order as that with the training materials at
every layer count, and the gap narrows as the layer count increases. The model thus designs with a material library it has never seen,
without retraining.
Figure~\ref{fgr:general}e,f shows representative designs.

\subsubsection{Beyond the training wavelength range}

We then tested bands beyond the 380--1400~nm training range. We took 140
stacks of 10--15 layers built from training materials whose optical constants
extend to 2500~nm, and used their spectra on nine bands on both sides of
1400~nm as targets (Note S7). The error increases beyond the training range
(Figure~\ref{fgr:general}g): relative to the error of the same stack inside
the training range, it is larger by a median factor of 1.4 in the
1400--1700~nm band and of 2.1 in the 1700--2160~nm band, the largest factor
among the nine bands. In every
band the median RMSE stays well below the standard deviation of the target
(Table S7), so the designs reproduce the spectral shape beyond the training
range with a moderate loss of accuracy. Figure~\ref{fgr:general}h,i shows one
14-layer stack evaluated on the two bands adjacent to the boundary.

\subsubsection{Idealized targets that cannot be realized exactly}

In all of the preceding tests, the target spectra were computed from actual stacks and were attainable by
construction. Real design targets are usually
idealized: no stack with a finite number of layers reproduces them exactly. We constructed 46 idealized targets covering common coating functions
(nine categories, listed in Note S8).

\begin{figure}[p]
  \centering
  \includegraphics[width=0.76\linewidth]{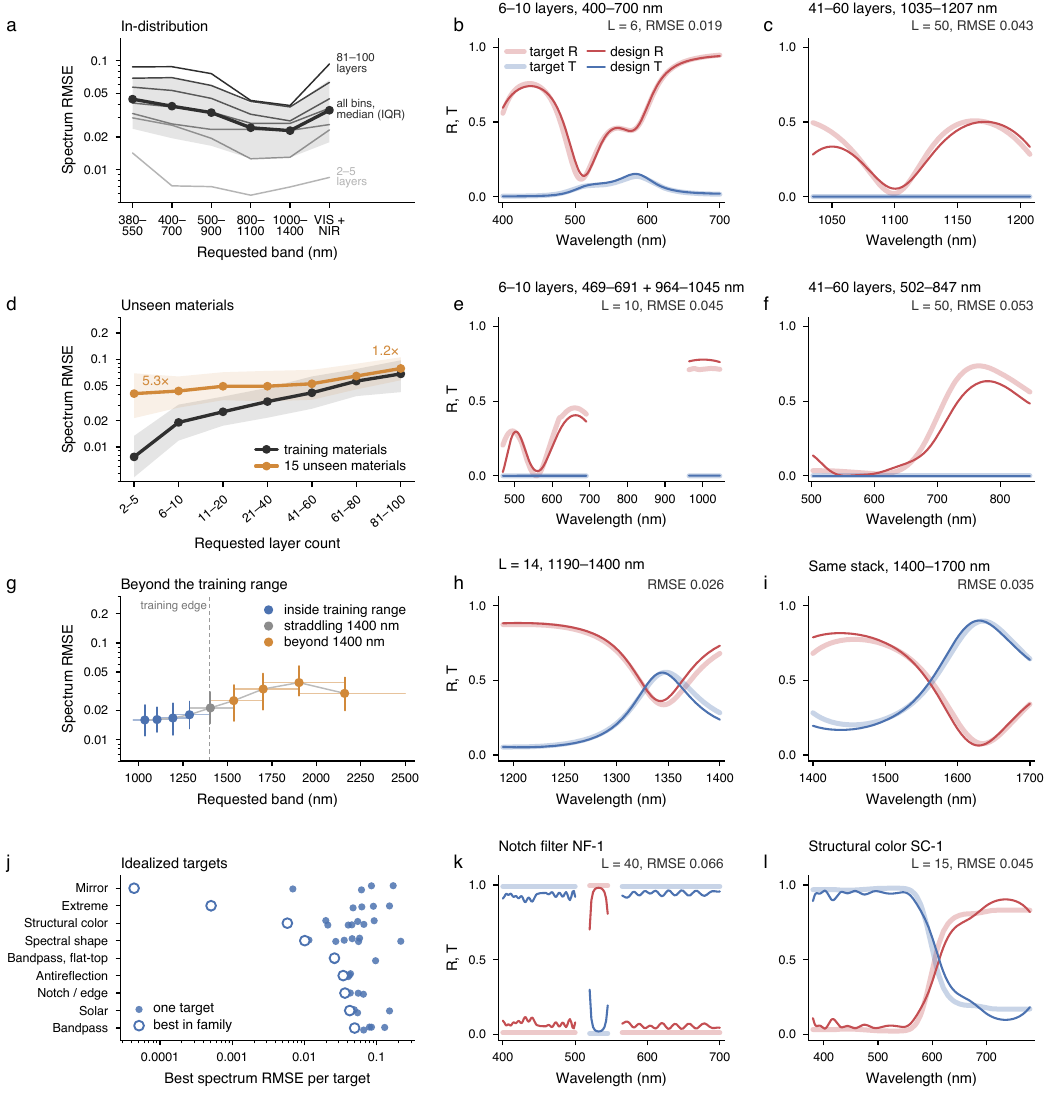}
  \caption{Performance of IrisFlow. (a) Inside the training distribution:
  RMSE versus the wavelength range within which the target band was drawn (black, median and interquartile range
  over all seven layer-count ranges; gray, individual ranges from 2--5 to
  81--100 layers, light to dark). (b, c) Representative designs, chosen close
  to the median of their layer-count range. (d) RMSE versus specified layer count for the training materials (black, in-distribution) and for 15
  held-out materials (orange); labels give the ratio of the two medians.
  (e, f) Representative designs with RMSE close to the median. (g) Bands beyond the training wavelength range: the same 140 stacks (10--15
  layers) evaluated on nine bands of equal width in $1/\lambda$. Horizontal bars
  mark the band extent and dots its $1/\lambda$ center; the vertical axis is RMSE (median and interquartile range). Blue, entirely inside the
  training range; gray, straddling the boundary; orange, entirely outside;
  the dashed line marks the 1400~nm limit of the training range. (h, i) Spectra of one 14-layer stack evaluated on the two bands adjacent to the boundary (1190--1400 and 1400--1700~nm).
  (j) Minimum RMSE for each of the 46 idealized targets. (k, l) Representative designs. In all spectral panels the thick
  line is the target and the thin line the re-simulated design; $R$ in red,
  $T$ in blue.}
  \label{fgr:general}
\end{figure}

Figure~\ref{fgr:general}j gives, for each target, the lowest RMSE among the
designs sampled at all layer counts; it ranges from $4.4\times10^{-5}$ to 0.22
across the 46 targets. A low error alone does not show that the model designs toward
the target, because it could also be reproducing a training spectrum that
happens to resemble it. To rule this out, we searched the training set for the spectrum
closest to each target on the target's own wavelength grid (Note S8). For 37 of the 46 targets, the best generated design is closer to the target
than this nearest training spectrum, so the model constructs structures that
the training set does not contain. Figure~\ref{fgr:general}k,l
shows representative results for a notch filter and a structural-color target.

\subsection{Comparison with classical optimization methods}

We compared IrisFlow with needle optimization with gradual
evolution,\cite{tikhonravov1996,tikhonravov2007} a genetic algorithm (GA),
particle swarm optimization (PSO), and random search on the 46 idealized
targets of Figure~\ref{fgr:general}j. All methods used the same materials,
thickness windows, stacking rules, and maximum layer count, and cost was counted
as the number of TMM-equivalent evaluations (Note S9). For each target, the
ten best designs sampled by IrisFlow were refined in thickness by
L-BFGS-B,\cite{byrd1995} and the best refined design was kept.

As shown in Figure~\ref{fgr:compare}a,b, IrisFlow with this refinement reaches a lower median RMSE than needle optimization, the strongest of the optimizers, with
about 500 times fewer evaluations, and is more accurate than GA, PSO, and
random search. When each optimizer is stopped at the cost that IrisFlow used,
IrisFlow gives the lower RMSE on every target. Per-target results and
statistical tests are given in Note S9.

\subsection{Comparison with OptoGPT}

We used the publicly released OptoGPT weights and compared the two models on
19 targets: the design targets shown in the OptoGPT paper, excluding the
validation spectra taken from its dataset, and the idealized targets in its
released code that satisfy $R + T \le 1$, keeping one of any targets that differ
only by a small shift of the band edges or of the reflectance
level.\cite{ma2024optogpt} IrisFlow (125.4 million
parameters) is larger than OptoGPT (about 58 million), so we also trained a smaller one to compare with OptoGPT. At inference, IrisFlow received the OptoGPT optical constants of 15
of the 18 materials in the OptoGPT library, because its candidate set is capped
at 15, whereas OptoGPT sampled from its full 18-material vocabulary. The two
models also use different thickness ranges, 5--300~nm for IrisFlow and
10--500~nm for OptoGPT. Each
model sampled 1000 designs for each target. The designs of both models were
re-simulated with the same TMM solver and optical constants on a common
wavelength grid, and for each model the lowest RMSE was reported (Note S10).

\begin{figure}[htbp]
  \centering
  \includegraphics[width=\linewidth]{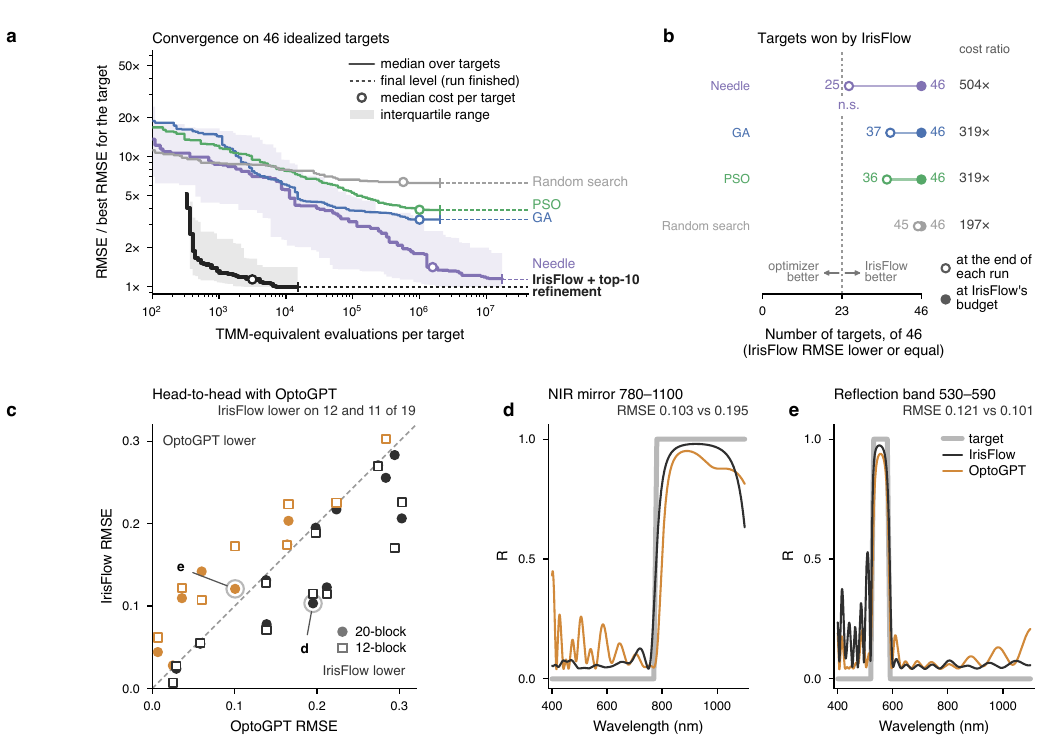}
  \caption{Comparison with existing methods. (a) Classical optimizers on the 46 idealized targets: median over targets of
  the RMSE of the best design found within a given number of TMM-equivalent
  evaluations, divided by the lowest RMSE that any method reached on that
  target; shading, interquartile range (IrisFlow and needle optimization). Lines turn dashed beyond the cost at which the runs on all targets have
  ended; open circles mark the median cost per target. IrisFlow + top-10
  refinement, 50 sampled designs per layer count followed by thickness
  refinement of the ten best; needle optimization, best of six starts; GA, PSO,
  and random search, median of three seeds. (b) Number of targets won by
  IrisFlow, that is, with an RMSE lower than or equal to the optimizer's (equal: relative
  difference below $10^{-5}$), at the end of each run (open) and with each optimizer
  stopped at IrisFlow's cost (filled); n.s., not significant (Wilcoxon
  signed-rank test, Holm correction); right, median ratio of the optimizer's
  cost to IrisFlow's (Note S9). (c) RMSE of the best IrisFlow
  design against that of the best OptoGPT design on 19 targets; circles,
  20-block model; squares, 12-block model; dark, IrisFlow lower; orange,
  OptoGPT lower; dashed line, equal RMSE. (d) A target on which IrisFlow performs better by the largest factor (near-infrared mirror, 780--1100~nm). (e) A target on which OptoGPT performs better (reflection band 530--590~nm). In d and e, only $R$ is shown. Thick gray lines
  are the targets.}
  \label{fgr:compare}
\end{figure}

As shown in Figure~\ref{fgr:compare}c, IrisFlow gives a lower RMSE than
OptoGPT on 12 of the 19 targets, with a median RMSE of 0.131 versus 0.164
(mean 0.146 versus 0.153), although the OptoGPT optical constants were
unseen by IrisFlow and were used to train OptoGPT; the smaller 12-block model
(79.1 million parameters) does so on 11. Figure~\ref{fgr:compare}d,e
shows a target on which each model performs better, and the per-target
comparison is given in Note S10.

\subsection{Design under constraints}

As shown in Figure~\ref{fgr:control}a, we imposed two kinds of local
constraints during sampling (Methods), without retraining, with the five
candidate materials of Figure~\ref{fgr:coat} on a target with reflectance 0.001
over 400--700~nm, reflectance 0.999 over 1000--1400~nm, and zero transmittance:
in partial-structure completion, some layers are fixed in material; in
fixed-sequence thickness inference, the whole material sequence is fixed and
only the thicknesses are generated. In both tasks, the best design reproduces
the target with an RMSE of 0.009 (Figure~\ref{fgr:control}b).

\subsection{Oblique incidence}

\begin{figure}[htbp]
  \centering
  \includegraphics[width=\linewidth]{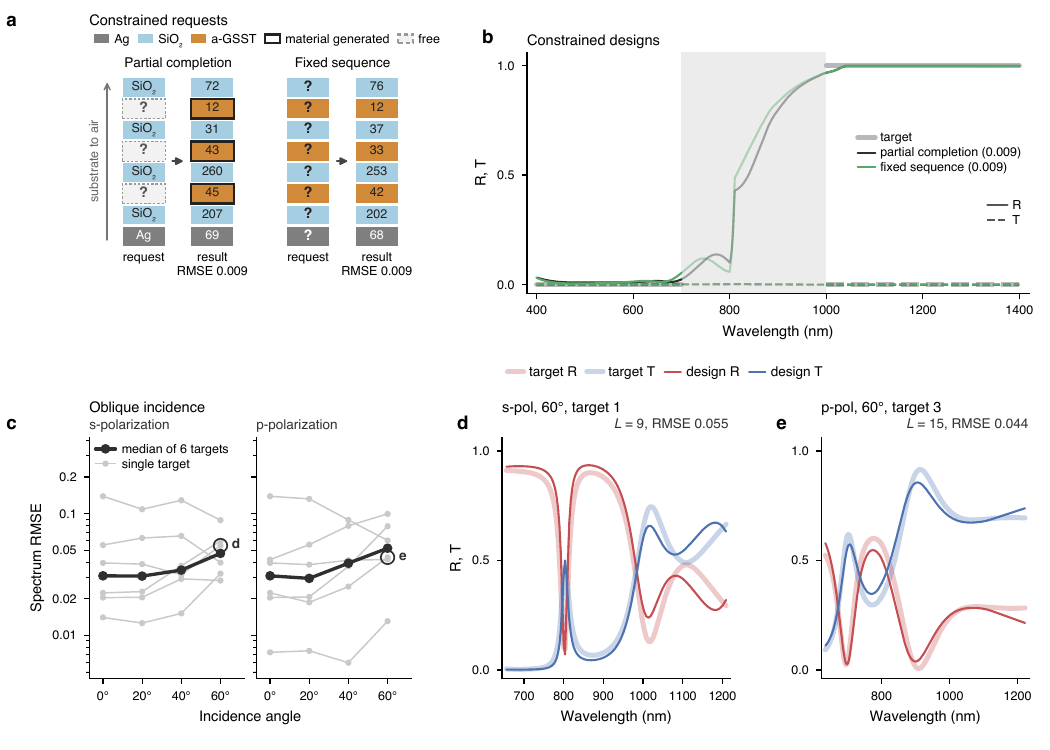}
  \caption{Controllable generation and oblique incidence. (a) Two constrained design tasks with the five candidate materials of Figure~\ref{fgr:coat}, each shown as a specification-and-result pair (substrate
  at the bottom, air at the top; layer heights not to scale). Left:
  partial-structure completion, in which the \ce{Ag} bottom layer and four
  \ce{SiO2} layers are fixed in material, while the materials of the three open
  slots (dashed boxes, ?) and all thicknesses are generated by the model. Right:
  fixed-sequence thickness inference, in which all eight materials are fixed
  and only the thicknesses (?) are generated. Bold outlines mark
  the layers whose material was chosen by the model; numbers are the
  generated thicknesses in nm. (b) Re-simulated
  spectra of the best design for each constrained design task ($N=500$); RMSE in
  parentheses; thick gray, target; solid, $R$; dashed, $T$; the gray band marks
  700--1000~nm, where the target imposes no requirement, and the design spectra
  are drawn lighter there. (c) Oblique incidence: RMSE at $0$--$60^\circ$ for six targets per polarization (left, s-polarization; right, p-polarization;
  gray, individual targets; black, median); circles mark the designs shown in
  d and e. (d, e) Representative $60^\circ$ designs for the two
  polarizations (thick line, target; thin line, design), chosen close to the
  median at that angle.}
  \label{fgr:control}
\end{figure}

All of the preceding tests were performed at normal incidence, whereas many
coatings are used at oblique incidence or for a specific polarization. Using
the conversion described in Methods, we supplied targets at angles of
incidence up to $60^\circ$ together with the correspondingly transformed
optical constants of the candidate materials, and evaluated each design by an
exact oblique-incidence TMM calculation with the original optical constants. The median error changes little
up to $20^\circ$ and rises at larger angles (Figure~\ref{fgr:control}c); at $60^\circ$ it is 1.5 times the
normal-incidence value for s-polarization and 1.7 times for p-polarization.
The difference between the two polarizations lies within the spread of the
six individual targets, although only the s-polarized conversion is exact at
the layer level (Note S11). The
conversion thus extends the model to oblique-incidence design without
retraining; Figure~\ref{fgr:control}d,e shows representative $60^\circ$
designs for the two polarizations.

\subsection{Design of colored solar-reflective coatings}

We used IrisFlow in a practical design workflow to design four colored
solar-reflective coatings for our deposition system; the coatings display
yellow, magenta, cyan, or black and reflect the solar near-infrared (targets in
Note S12). The candidates were the five materials available in this system: \ce{Ag}, \ce{SiO2}, \ce{Ti}, amorphous \ce{Ge2Sb2Se4Te1} (GSST), and
\ce{LaTiO3}. We calibrated the optical constants of each material in the
visible range with a spectrophotometer; they agree with the training values
except those of \ce{GSST} (Note S6). Designs with an \ce{Ag} layer next to a \ce{GSST} layer were excluded,
because \ce{Ag} reacts with the chalcogen constituents of \ce{GSST} (Se and Te)
at a direct interface.

\begin{figure}[htbp]
  \centering
  \includegraphics[width=\linewidth]{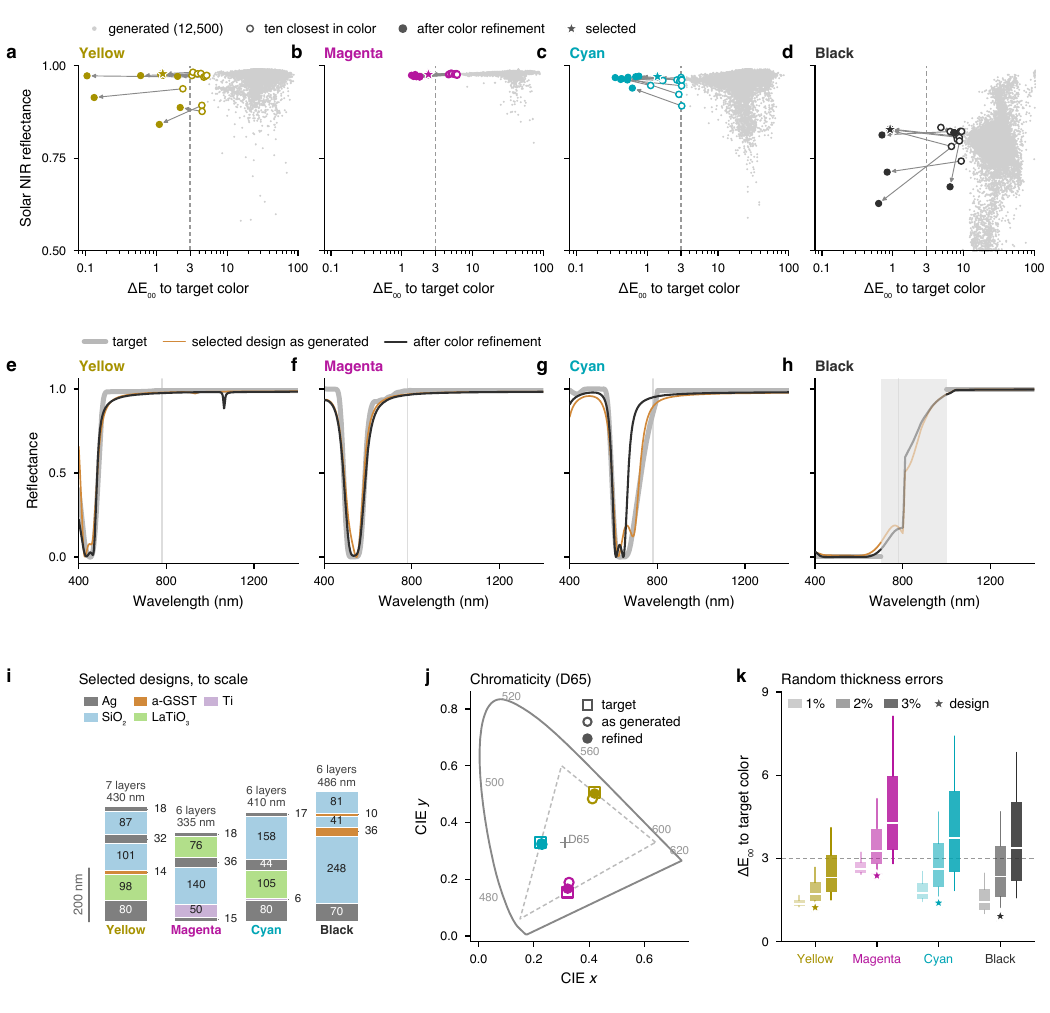}
  \caption{Design of four colored solar-reflective coatings. (a--d)
  AM1.5G-weighted reflectance over 780--1400~nm versus color difference \dE{}
  from the target color; gray, the 12,500 designs sampled for each color; open
  and filled circles, the ten designs closest to the target color before and
  after thickness refinement; star, the selected design; dashed line,
  $\dE=3$. (e--h) Reflectance of the selected designs; thick gray, target;
  orange, as generated; black, after refinement; the gray band in h marks
  700--1000~nm, where the black target imposes no requirement, and the design
  spectra are drawn lighter there; the thin vertical line marks 780~nm. (i)
  Layer structures of the selected designs (to scale in thickness,
  fused-silica substrate at the bottom; numbers are thicknesses in nm). (j) CIE
  1931 chromaticity (D65); squares, the three chromatic targets; open and filled
  circles, the selected designs as generated and after refinement; dashed
  triangle, sRGB gamut. (k) \dE{} from the target color under random relative
  thickness errors with a standard deviation of 1, 2, and 3\% in every layer,
  4000 perturbed stacks per level; boxes, interquartile range; whiskers, 10th to
  90th percentile; stars, the selected designs without thickness errors.}
  \label{fgr:coat}
\end{figure}

We used IrisFlow to sample 12,500 designs for each color. As shown in
Figure~\ref{fgr:coat}a--d, the designs spread over a wide range of color difference \dE{} from the target and of
solar near-infrared reflectance, since IrisFlow is conditioned on the target
spectrum, and designs with similar spectral errors can
differ widely in color. The ten designs closest to the target
color were therefore refined in thickness, with their materials fixed, to
minimize \dE{} (Note S12). After refinement, all ten designs of each chromatic
color and six of the black ones lie within $\dE=3$. Among them, the designs
within 0.005 of the highest solar near-infrared reflectance were kept, and the
one with the fewest layers, which is the easiest to deposit, was selected for
each color; the refinement changed the solar near-infrared reflectance of the
selected designs by at most 0.022 (Figure~\ref{fgr:coat}e--j; Table S14).

Every layer thickness of the selected designs was then perturbed by a random
relative error (Figure~\ref{fgr:coat}k). For an error of 2\% per layer, the median \dE{} from the target stays
below 3 for yellow, cyan, and black and is 3.3 for magenta, and the solar
near-infrared reflectance changes by less than 0.01 in 90\% of the perturbed
stacks (Note S12).

\section{Conclusions}

In this work, we introduce IrisFlow. It couples discrete material generation
and continuous thickness generation in a single generative model, which
receives the candidate materials as their optical constants and the target
spectrum together with its wavelengths. Because a design task is specified by
these inputs, one trained model designs with material libraries it has never
seen, in bands beyond its training range, and at oblique incidence, and it
completes partially fixed structures or generates thicknesses for a fixed
material sequence, all without retraining. After a short local refinement it
reaches an accuracy comparable to that of needle optimization with far fewer
transfer-matrix evaluations, and it achieves lower errors than OptoGPT on most
targets. With the calibrated optical constants of an
existing deposition system, we use it to design colored solar-reflective coatings that
reach their target colors.

The formulation should carry over to other inverse-design problems that
combine discrete choices with continuous parameters, such as metasurface or
photonic-crystal design, where a material must be chosen for each element
together with its dimensions.

\begin{suppinfo}
Details of the discrete--continuous flow matching and the sampler;
training-data generation, network structure, and training settings; optical
constants of the training, held-out, and calibrated materials; and the
protocols and per-target results of all tests, including the comparisons with
classical optimizers and OptoGPT, the oblique-incidence conversion, and the
design of the colored coatings.
\end{suppinfo}

\begin{acknowledgement}
This work was supported by the National Natural Science Foundation of China
(Grant No. 62375231), the Fujian Province Science and Technology Planning
Project of China (Grant No. 2022H6015), and the Xiamen University President's
Fund Project of China (Grant No. 20720242021).
\end{acknowledgement}

\section*{Data and Code Availability}
The code for data generation, training, and design, the trained model
weights, the training and test data, the optical-constant libraries of the
training and held-out materials, and the data behind Figures 3--6 are publicly
available at \url{https://github.com/z-one-l/IrisFlow}.

\bibliography{IrisFlow}

\clearpage
\includepdf[pages=-]{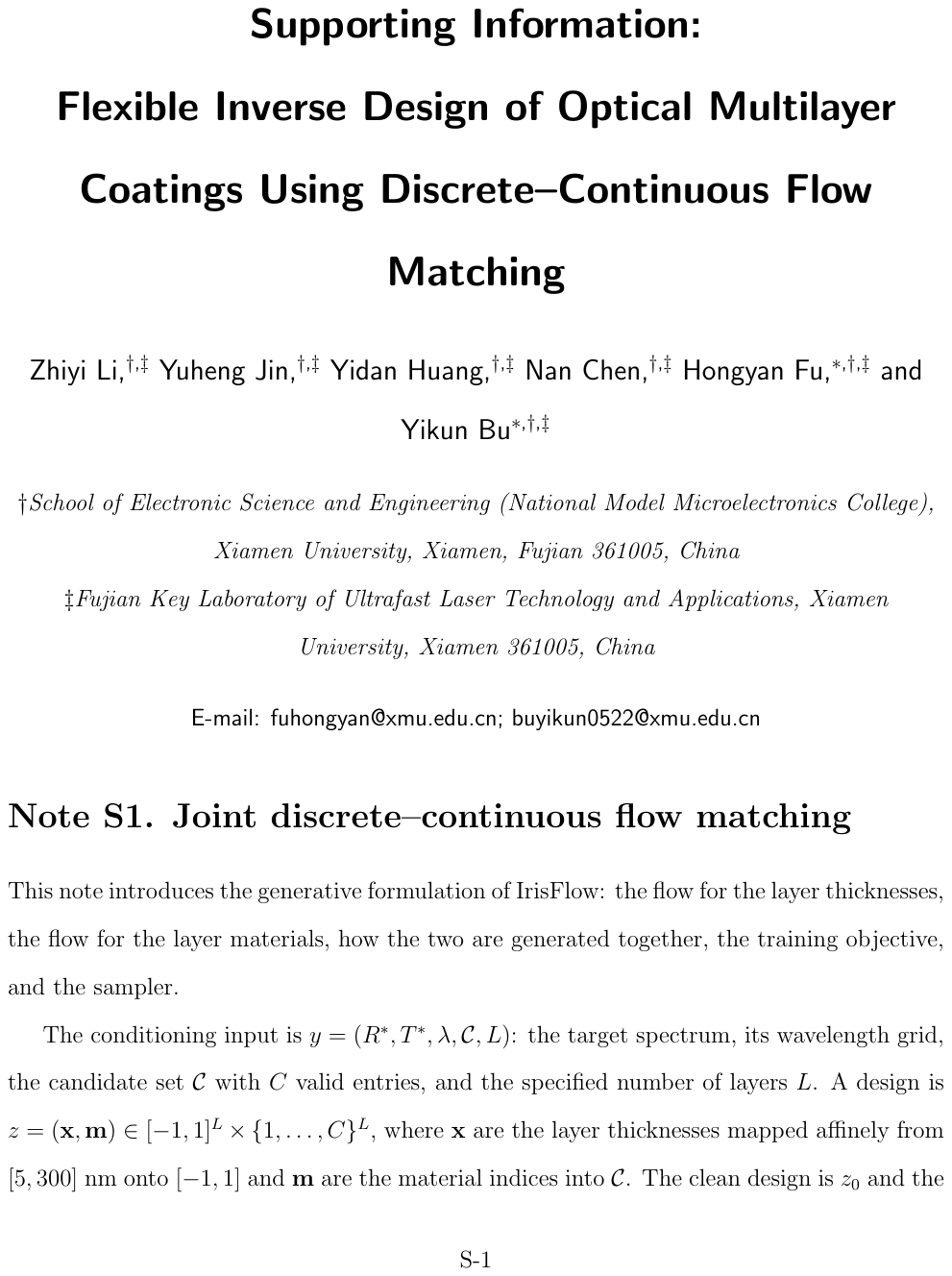}

\end{document}